# Nitride-based interfacial layers for monolithic tandem integration of new solar energy materials on Si: The case of CZTS

Filipe Martinho[1,*,†], Alireza Hajijafarassar[2,*,†], Simón Lopez-Marino[2], Moises Espíndola-Rodríguez[3], Sara Engberg[1], Mungunshagai Gansukh[1], Fredrik Stulen[4], Sigbjørn Grini[4], Stela Canulescu[1,*], Eugen Stamate[2], Andrea Crovetto[5], Lasse Vines[4], Jørgen Schou[1], Ole Hansen[2]

[1]Department of Photonics Engineering, Technical University of Denmark, DK-4000 Roskilde, Denmark.

[2]DTU Nanolab, Technical University of Denmark, DK-2800 Kgs. Lyngby, Denmark

[3]DTU Energy, Technical University of Denmark, DK-4000 Roskilde, Denmark.

[4]Department of Physics, University of Oslo, 0371 Oslo, Norway

[5]DTU Physics, Technical University of Denmark, DK-2800 Kgs. Lyngby, Denmark.

**Corresponding Author**

*Email: filim@fotonik.dtu.dk (Filipe Martinho), alhaj@dtu.dk (Alireza Hajijafarassar), stec@fotonik.dtu.dk (Stela Canulescu)

† - These authors contributed equally to this work

**Abstract**

The monolithic tandem integration of third-generation solar energy materials on silicon holds great promise for photoelectrochemistry and photovoltaics. However, this can be challenging when it involves high-temperature reactive processes, which would risk damaging the Si bottom cell. One such case is the high-temperature sulfurization/selenization in thin film chalcogenide solar cells, of which the kesterite $Cu_2ZnSnS_4$ (CZTS) is an example. Here, by using very thin (<10 nm) TiN-based diffusion barriers at the interface, with different composition and properties, we demonstrate on a device level that the protection of the Si bottom cell is largely dependent on the barrier layer engineering. Several monolithic CZTS/Si tandem solar cells with open-circuit voltages ($V_{oc}$) up to 1.06 V and efficiencies up to 3.9% are achieved, indicating a performance comparable to conventional interfacial layers based on transparent conductive oxides, and pointing to a promising alternative design in solar energy conversion devices.

**Introduction**

The prospect of fabricating a monolithically integrated two-terminal (MI-2T) tandem device for solar energy conversion has attracted considerable interest, due to the possibility of higher conversion efficiencies and to inherent functional advantages in the fields of both photovoltaics and photoelectrochemistry.[1] In photovoltaics, a MI-2T tandem solar cell minimizes the number of processing steps and interconnections of 2T tandem configurations, making it the most suitable configuration for large-scale industrialization, in particular when leveraging the currently existing technology for crystalline Si (c-Si).[2] As a result, research for new materials as partner with Si is ongoing, along with new tandem design concepts. Regarding the latter, we have recently proposed the use of a thermally resilient Tunnel Oxide Passivated Contact (TOPCon) Si cell structure in combination with nitride-based diffusion barriers for the monolithic integration of materials synthesized under extreme conditions, such as high-temperature sulfurization of the thin-film chalcogenide $Cu_2ZnSnS_4$ (CZTS).[3] In order to achieve functional CZTS/Si monolithic devices, it has been suggested that strategies for protecting the bottom Si cell need to be developed.[3–5] In this work, we approach this problem through a comparative study of CZTS/Si tandem cells fabricated using three different types of TiN-based diffusion barrier layers. For the first and second, we use atomic layer deposition (ALD) to produce 5 and 10 nm TiN barrier layers, and a 10 nm $TiO_xN_y$ barrier layer. The difference in oxygen content was achieved by running the TiN ALD recipe without prior chamber passivation (i.e. a sequence of dummy TiN depositions), thereby allowing a higher oxygen background level. We have previously demonstrated that this procedure increases the transparency of the TiN layer and leads to a promising diffusion barrier quality.[3] Furthermore, metallization studies on Si have revealed that different oxygen contents of TiN films can result in significant changes in the barrier performance against the diffusion of Cu and Al.[6–8] For the third barrier, we use a sputtered TiN layer modified by an intermediate Al layer, in a configuration TiN (5 nm)/Al (2 nm)/TiN (5 nm). This configuration is known to improve the barrier resilience against Cu diffusion when a pre-annealing in air is used to segregate Al to the TiN grain boundaries. Due to the air annealing, $Al_2O_3$ stuffs the grain boundaries, reducing the grain boundary diffusion through the columnar structure of thin TiN films.[9]

Through this work, it is shown on a device level how the different barrier layers tested achieve different degrees of success in regards to the compromise between diffusion barrier quality, optical transparency and electrical interconnection between top and bottom cell. As a result, CZTS/Si solar cells with efficiencies up to 3.9% and $V_{oc}$ up to 1.06 V are achieved. Notably, the CZTS/Si tandem cell performance obtained in this work is comparable to that obtained using conventional interface layers based on transparent conductive oxides (TCOs)[4], suggesting that this diffusion barrier approach could be a promising alternative tandem device configuration. If successful, one potential advantage of this approach is the possibility of tuning the barrier layer properties according to the top cell material and synthesis conditions, in order to explore the monolithic integration of new materials on Si.

**Results and Discussion**

The effectiveness of the three TiN-based interfacial layers was evaluated by a combination of measurements of the effective minority carrier lifetime on the bottom Si cell (hereafter referred to as lifetime), the tandem J-V characteristics and their respective external quantum efficiency (EQE), and the optical transmittance of the TiN layers. **Figure 1** shows a comparison of the Si lifetime after CZTS processing of three tandem cells using TiN layers with different thicknesses (5 and 10 nm, denoted as TiN-5 and TiN-10) and oxygen contents (TiON-10). Here, CZTS processing refers to co-sputtering of Cu, ZnS and SnS precursors and annealing in a sulfur atmosphere at 575 °C. The initial lifetime of all three wafers (before CZTS processing) was 1 ms. Furthermore, the two tandem pieces shown in **Figure 1 (a)** have an additional SiN surface patterning confining the Si exposure to CZTS to squares with 3×3 mm$^2$, as illustrated by the accompanying schematic drawing in **Figure 1 (a)**. The lower lifetimes in the small squares (<

0.3 ms) clearly highlight the impact of CZTS fabrication on the lifetime of the Si bottom cell. We note that the thickness of the TiN barrier layer had a negligible effect on the Si protection, which was also confirmed in non-patterned versions of this experiment, shown in **Figure S1** of the supporting information (SI). From our previous work[3], this reduction in lifetime can be connected to contamination of the Si bottom cell with elements from CZTS, and to the loss in surface passivation quality of the Si front and back surfaces. However, **Figure 1 (b)** shows that the lifetime is clearly uniformly higher across the wafer when a barrier layer with a higher oxygen content is used. Notably, the lifetime is nearly the same as the initial, prior to CZTS processing.

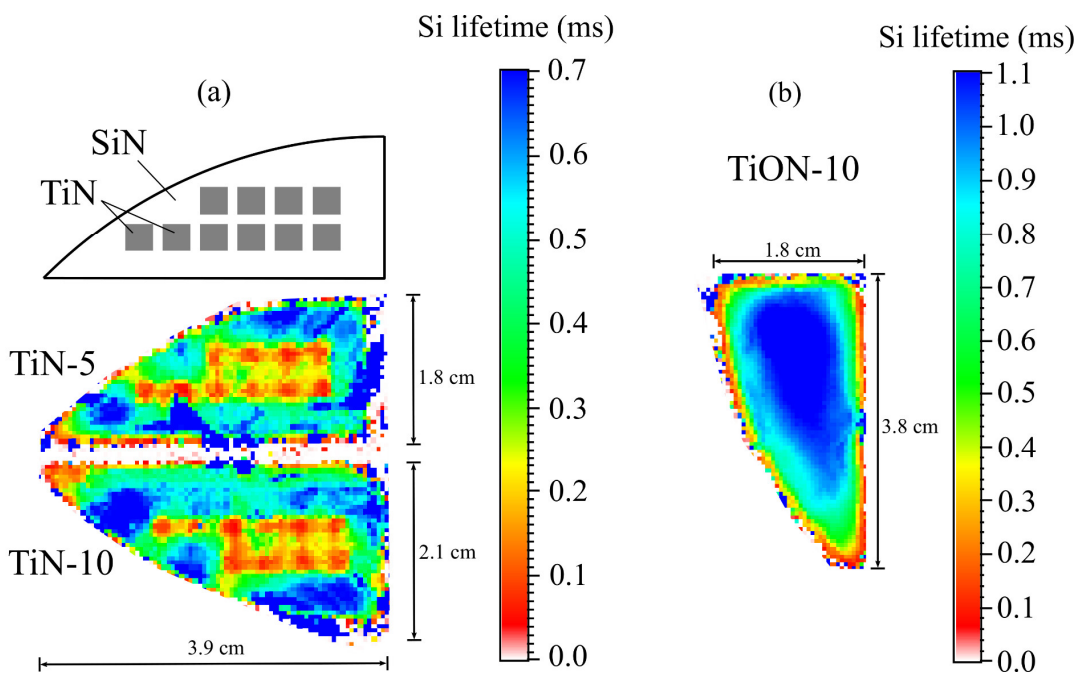

**Figure 1** – Minority carrier lifetime of the Si bottom cell after CZTS fabrication: (a) Two SiN-patterned wafers with a TiN barrier of 10 nm (below) and 5 nm (middle). The schematic drawing (above) shows the patterning, which defines 10 square windows where CZTS directly contacts

with TiN/Si. The square windows show a clearly degraded lifetime; (b) A non-patterned wafer with a TiON barrier of 10 nm.

The chemical composition of the TiN-10 and TiON-10 layers was determined using X-ray Photoelectron Spectroscopy (XPS), and is shown in **Figure 2** and **Table 1**. **Figure 2 (a)** and **(b)** show that the Ti 2p peaks could be deconvoluted into three main sets of Ti 2p spin-orbit doublets (Ti $2p_{3/2}$ and Ti $2p_{1/2}$ peaks) each separated by 6 ± 0.2 eV. The components were identified as TiN (Ti $2p_{3/2}$, 454.9 eV)[10–12], $TiO_2$ (Ti $2p_{3/2}$, 458.2 eV)[12] and an intermediate oxynitride compound $TiO_xN_y$ (Ti $2p_{3/2}$, 455.6 eV).[12] To achieve a better fit for the TiN-10 sample, an additional pair of peaks (at 457 eV and 462.6 eV) were added to account for the satellite features of the pure TiN phase, which are known as shake-up events.[10,11] The satellite components were neglected for the TiON-10 sample as the amount of the pure TiN phases is much lower compared to that of the TiN-10 sample. Moreover, the asymmetric nature of both N 1s and O 1s transitions (see **Figure 2** (c) and (d)) further confirms the co-existence of nitride, oxide, and oxynitride phases. By comparing the intensities of the Ti $2p_{3/2}$, N 1s, and O 1s peaks, the composition of the films was estimated, as summarized in **Table 1**. The TiON-10 barrier layer had 37 at% oxygen, resulting from the co-existence of TiN, $TiO_2$ and $TiO_xN_y$ phases. In comparison, the TiN-5 and TiN-10 barriers, where no oxygen was intentionally introduced, still exhibited residual oxygen around 13 at%, from the less-pronounced $TiO_xN_y$ and $TiO_2$ signatures. However, it is well-known that it is notably difficult to achieve low oxygen levels in nitride films, as the oxidation is thermodynamically favorable (with a negative Gibbs free energy variation), and oxygen contents up to 20% can occur near the surface[12]. While no improvement in Cu barrier properties is expected for oxygen contents up to 15%[7], it has been reported that for higher oxygen contents a stuffing effect occurs, resulting in improved barrier properties against Cu diffusion[8], which is in line with our results.

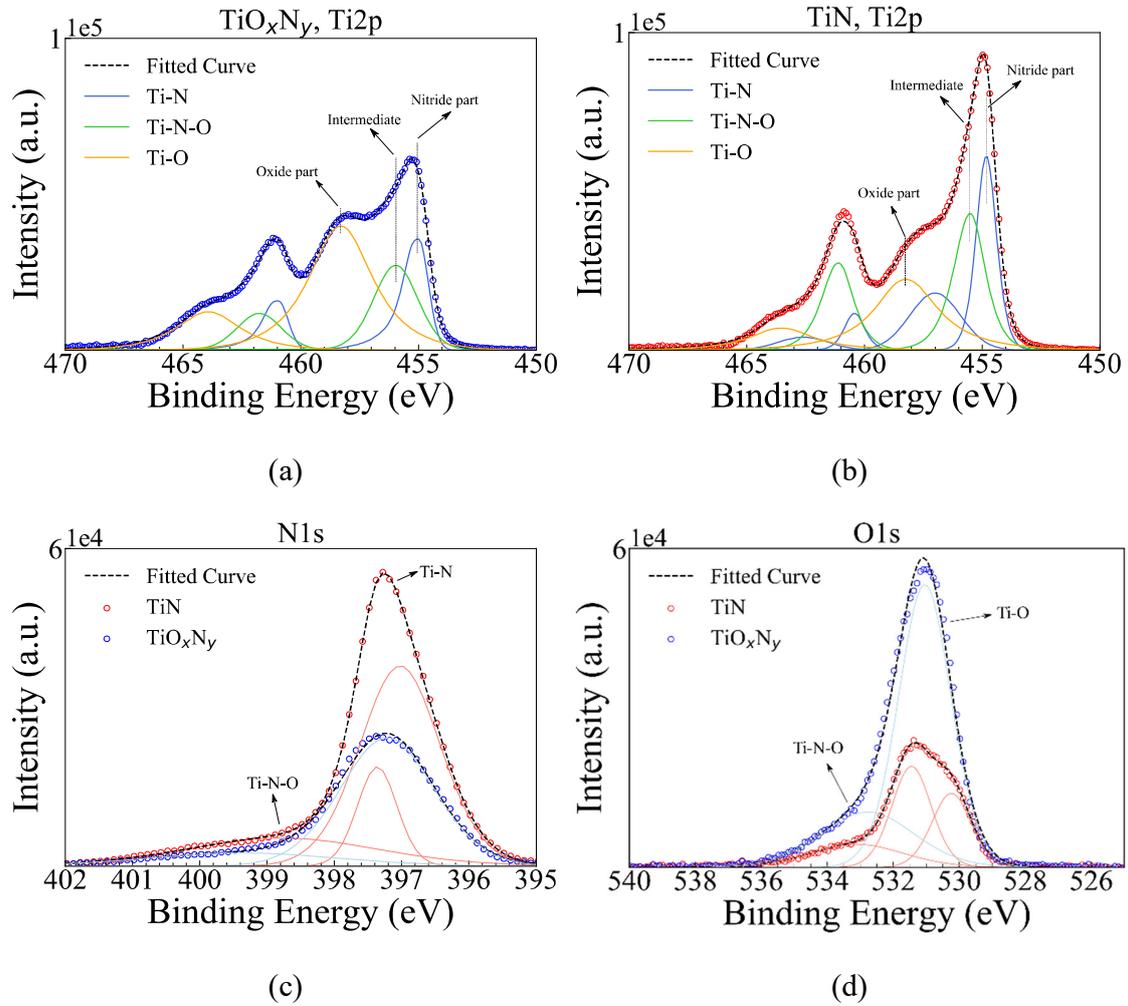

**Figure 2** – XPS spectra for TiN-10 and TiON-10 samples. (a) and (b) Ti 2p spectrum of TiON-10 (blue circles) *and* TiN-10 (red circles) and their corresponding N 1s (c) and O 1s (d) spectra after Shirley background subtraction. The dashed line represents the envelope of all individual components. The satellite (shake-up) components are neglected for the TiON-10 sample.

**Table 1** – Atomic composition (in atomic percentage) for the 10 nm TiON and TiN barriers. The composition was estimated using the Ti $2p_{3/2}$, O 1s and N 1s peaks.

| Barrier layer | Ti (at. %) | O (at. %) | N (at. %) |
|---|---|---|---|
| **TiON-10** | 40 | 37 | 24 |
| **TiN-10** | 47 | 13 | 40 |

Using Secondary Ion Mass Spectrometry (SIMS), we have further confirmed the superior diffusion barrier layer properties of TiON-10 by conducting a standard Cu-diffusion test on device wafers similar to the ones used for tandem cell fabrication. For this purpose, metallic Cu was deposited on the Si device wafer, protected either by TiN (10 nm) or TiON (10 nm), and then annealed at 550 °C in vacuum for 15, 30, 45 and 60 minutes, as described in previous work.[3] The corresponding SIMS results are shown in **Figure 3**. The comparison between the two barrier layers reveals a clear one order of magnitude difference in Cu concentration between the TiN-10 (solid lines) and TiON (dashed lines), demonstrating the superior barrier quality of the TiON-10 layer. Unfortunately, the high oxygen in the TiON layer also increases its resistivity and causes severe current blocking, causing an S-shape (or rollover) in the J-V characteristic, as we have found previously.[3]

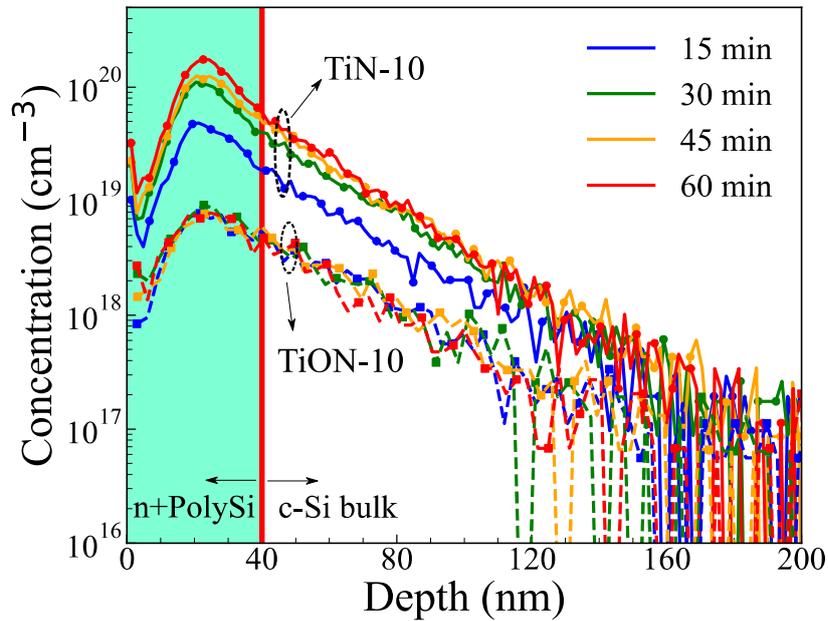

**Figure 3** – Quantitative Cu depth profile of the Si cell surface structures (n+PolySi contact, SiO$_2$ and bulk n-Si). The solid lines (circle markers) correspond to the TiN-10 barrier, and the dashed lines (square markers) to the TiON-10 barrier.

The light J-V characteristics and EQE spectra of two different tandem cells with the TiN-5 barrier layer are shown in **Figure 4 (a)** and **(b)**, respectively. In one case, the Si bottom cell has been compromised, corresponding to a barrier failure and severe degradation of lifetime, and in the other case the Si bottom cell was (at least partially) protected, retaining some of the initial lifetime after CZTS processing. This degradation leads to an overall drop in EQE, as can be seen for the Si bottom cell comparisons in **Figure 4 (b)**. As a result, a much lower current is extracted on the Si side, which limits the overall tandem cell current. This effect was also seen for the TiN-10 barrier, which is detailed in the SI (**Figure S2**). As mentioned above, the decrease in lifetime and performance can be explained by a combination of loss of surface passivation in the bottom Si cell (increase in surface recombination velocity) and a contamination of the Si bulk with elements from CZTS.

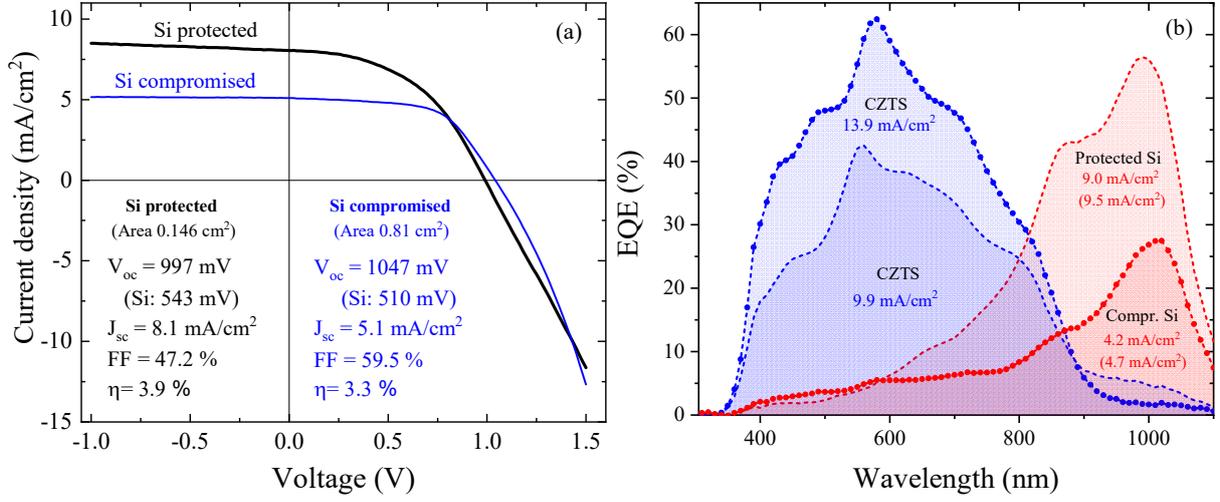

**Figure 4** – (a) J-V characteristic curve for a CZTS/TiN (5nm)/Si tandem with a compromised Si (blue) and a partially protected Si (black). The $V_{oc}$ of the Si bottom cell of the tandem is included in parenthesis; (b) EQE comparison of CZTS/Si tandem cells when the Si bottom cell is partially protected and when it is compromised.

To further investigate the influence of bulk and surface recombination in our results, we have derived a simple theoretical model for the Internal Quantum Efficiency (IQE). In this model, we consider a bottom cell with thickness $L$ with a perfectly collecting junction at the back, a certain finite bulk lifetime $\tau_{\text{bulk}}$ and a surface recombination velocity $S$ at the illuminated surface. By using Donolato's reciprocity theorem for the collection efficiency[13,14] and solving the appropriate minority carrier diffusion equation with a matching flux condition on the non-ideal surface, the IQE can be shown to be

$$\text{IQE} = \frac{\alpha L_p}{2} \frac{\left(\frac{SL_p}{D} + 1\right)\frac{e^{-\alpha L}e^{\frac{L}{L_p}} - 1}{1 - \alpha L_p} + \left(\frac{SL_p}{D} - 1\right)\frac{e^{-\alpha L}e^{-\frac{L}{L_p}} - 1}{1 + \alpha L_p}}{\left(\cosh\left(\frac{L}{L_p}\right) + \frac{SL_p}{D}\sinh\left(\frac{L}{L_p}\right)\right)} \quad (1)$$

where $\alpha$ is the absorption coefficient, $D$ is the diffusion coefficient and $L_p$ is the diffusion length, with $L_p = \sqrt{D\tau_{bulk}}$. More details on the derivation of this model can be found in the supporting information. The surface recombination velocity $S$ and the bulk lifetime $\tau_{bulk}$ can be related to the experimentally measured effective minority carrier lifetime $\tau_{eff}$ using the expression[15,16]

$$S = \sqrt{D\left(\frac{1}{\tau_{eff}} - \frac{1}{\tau_{bulk}}\right)} \tan\left[\frac{L}{2}\sqrt{\frac{1}{D}\left(\frac{1}{\tau_{eff}} - \frac{1}{\tau_{bulk}}\right)}\right] \qquad (2)$$

Using **Equation (1)** and **Equation (2)**, we can then simulate IQE curves for silicon with some experimentally measured $\tau_{eff}$ and find possible matching combinations of $S$ and $\tau_{bulk}$. **Figure 5** shows simulated IQE curves for three different effective lifetimes of 50, 30 and 10 μs. These effective lifetimes were chosen because we systematically found our compromised samples to have experimental effective lifetime values in the range 10-50 μs (see **Figure 1 (a), Figure S1** and **Figure S6**). The set of $\tau_{bulk}$ and $S$ values matching $\tau_{eff}$ are listed as inset in each graph in **Figure 5**, from a dominant bulk recombination (low $S$) to a dominant surface recombination (high $S$) scenario. For the simulations, we used an $\alpha(\lambda)$ for Si as measured by Green and Keevers[17], a thickness of $L = 350$ μm, and the diffusion coefficient for holes in Si of $D = 12 \text{ cm}^2\text{s}^{-1}$.

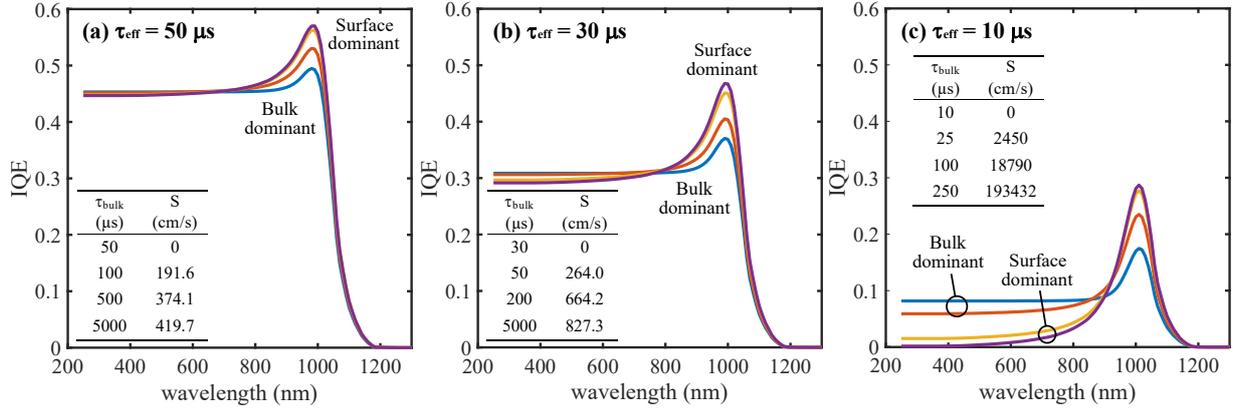

**Figure 5** – Simulated internal quantum efficiency curves for Si with effective minority carrier lifetimes ($\tau_{\text{eff}}$) of (a) 50 µs, (b) 30 µs and (c) 10 µs. The inset tables show possible combinations of bulk lifetime $\tau_{\text{bulk}}$ and surface recombination velocity $S$ that match $\tau_{\text{eff}}$, from dominant bulk recombination (low $S$) to dominant surface recombination (high $S$).

The expression for $S$ in **Equation (2)** was used because it is still a good approximation for large values of S, unlike simpler expressions.[15] However, note that the tangent function has an argument limit of $\pi/2$, which places constraints on the values of $\tau_{\text{eff}}$ and $\tau_{\text{bulk}}$. For instance, for very large $\tau_{\text{bulk}}$ the expression is only valid if $\tau_{\text{eff}} > 10$ µs, given our $D$ and $L$ values. As mentioned above, our experimental $\tau_{\text{eff}}$ values are above this limit.

The effect of surface recombination produces a distinctive peak in the long wavelength region (near 1000 nm, as shown in **Figure 5 (a), (b)** and **(c)**), which appears to match the shape of our experimental EQE curves (see also **Figure S2** and **Figure S7**). Therefore, we assign the loss of lifetime in our compromised samples primarily to a dominant surface recombination due to depassivation of the polySi/SiO$_2$/Si interfaces. Furthermore, we have investigated in previous work the possibility of passivating the Si surface only after CZTS annealing and etching (end-passivation), thereby avoiding the depassivation effect.[3] It was found that end-passivated samples exhibited lifetimes above 1 ms, further indicating that surface recombination should be the dominant effect in this work, where the CZTS annealing affects the pre-existing polySi/SiO$_2$ surface passivation. Nevertheless, we note that some bulk degradation could still occur, as indicated by some bulk contamination measured by SIMS in this work and in previous work.[3] In any case, a successful barrier layer will simultaneously minimize the impact of both bulk and surface degradation. Another important feature revealed by the simulations of **Figure 5** is that in compromised Si samples, where the effective lifetime is in the µs range, small variations in $\tau_{\text{eff}}$ produce large variations in the corresponding IQE spectra. This is possibly the reason for the clear difference in EQE between the *Si protected* and *Si compromised* samples in **Figure 4 (b)**. Even though the experimentally measured $\tau_{\text{eff}}$ for both samples is below 300 µs, small differences due to processing variabilities and variations in lifetime near the patterned TiN windows (which can clearly be seen in **Figure 1 (a)** for both the TiN-5 and TiN-10 cases) will appear as large differences in the respective IQE spectra. This also indicates that even the case labelled as *Si protected* represents at best only a partial protection. This is will be further discussed next.

The degradation of the bottom cell also results in $V_{\text{oc}}$ losses, although with a smaller impact on solar cell efficiency when compared to the short-circuit current ($J_{\text{sc}}$) losses. The $V_{\text{oc}}$ of the Si bottom cell of the tandems presented in **Figure 4** was measured by removing a small area of the top cell layers and contacting it with the negative probe. These values were 543 mV for the *Si protected* bottom cell and 510 mV for the *Si compromised* bottom cell. We note also in **Figure 4 (a)** that the total tandem cell $V_{\text{oc}}$ with the compromised Si is about 50 mV higher than its protected counterpart. However, several other effects contribute to this difference. One such effect is variability in CZTS performance – as shown in **Figure 4 (b)**, which we ascribe to baseline variability, even though the CZTS synthesis conditions were kept as constant as possible throughout these experiments. Furthermore, since the active area of the tandem cells was defined by cleaving the Si wafer, another important factor is shunting due to the exposed Si edges (which have a higher surface recombination velocity). Uncleaved wafers exhibited $V_{oc}$ values systematically higher than shown here, up to 1064 mV, as detailed in the SI (**Figure S3**).

It is also interesting to compare the results of **Figure 4** with the case of the TiON-10 barrier. This TiON-10 barrier was characterized in detail in our previous work and was used to achieve our first monolithic integration of CZTS on Si.[3] Then, a complete tandem cell was produced without any measurable degradation of the Si bottom cell due to CZTS fabrication, with post-fabrication carrier lifetimes well above 1 ms in the Si bulk. In this work, we have repeated this previous experiment in order to have comparable measurements of the Si bottom cell $V_{oc}$, and found a $V_{oc}$ of 661 mV for the bottom cell in the TiON-10 case. This value is significantly higher than for the TiN-5 and TiN-10 samples (shown below in **Table 2**), and is approximately the expected value of our in-house single junction Si cell when the light spectrum is filtered according to our CZTS bandgap (1.45 eV, as determined from the inflexion point on the EQE spectra). Note that all the bottom cell $V_{oc}$ values measured in this work are obtained from working tandem cells, with the top cell filtering out the lower wavelength part of the solar spectrum. This confirms that, as mentioned with the simulations above, the case labelled *Si protected* in **Figure 4 (a)** is only at best a partial protection, thus proving on a device level that only the TiON barrier seems to be completely successful in protecting the bottom Si cell. Unfortunately, it exhibits an extremely high resistivity (around 40 Ωcm), with a severe current-blocking behavior, meaning that it cannot be used in a functional tandem device. The TiN-5 and TiN-10 sacrifice some of the barrier quality but in turn provide a better electrical interconnection between bottom and top cell. Remarkably, despite the admittedly nonideal degradation of the bottom Si cell, our results are an improvement over the best CZTS/Si tandem cell achieved so far,[4] which further highlights the potential of the alternative tandem configuration proposed in this work. It is critical, then, to search for a better compromise between the barrier function and the interconnection function.

In order to further explore this possibility, we investigated a modified TiN barrier layer, in which a thin Al layer is sputtered in-between two TiN sputtering runs, in a TiN (5 nm)-Al (2 nm)-TiN (5 nm) configuration, hereafter labelled TiN-Al-10. The purpose was to achieve a similar stuffing effect observed for the TiON-10 barrier by stuffing the TiN grain boundaries with $Al_2O_3$ upon annealing in air[9], prior to CZTS fabrication. The J-V and EQE results are summarized in **Figure 6**. The results show that there is an improvement in the Si EQE compared to the *Si compromised* case shown in **Figure 4**, suggesting some degree of improvement in the barrier layer properties. However, the current was still far from the *Si protected* case. As for the $V_{oc}$, it was comparatively lower, and we systematically observed bottom cell $V_{oc}$s of around 400-450 mV for samples using this TiN-Al-10 barrier layer.

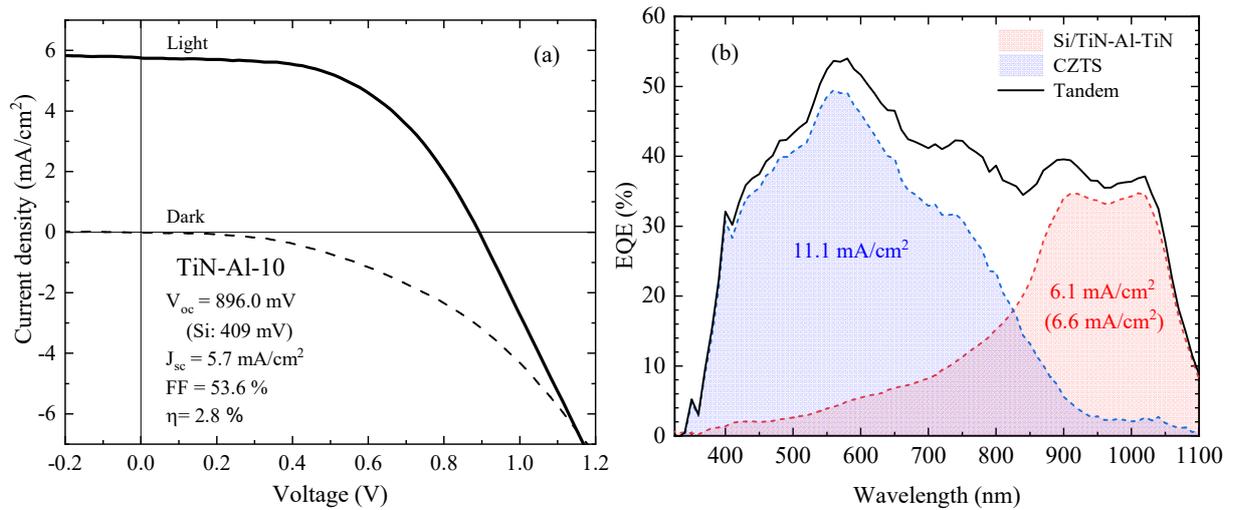

**Figure 6** – (a) Light and dark J-V characteristic curves for a tandem cell using a TiN (5 nm)-Al (2 nm)-TiN (5nm) barrier layer (TiN-Al-10 for short). The $V_{oc}$ of the Si bottom cell of the tandem is included in parenthesis; (b) Respective EQE spectra of the tandem cell.

These results can be understood by comparing the respective post-processing lifetimes, as was done previously in **Figure 1** for the TiN-5 and TiN-10 cases. In **Figure 7**, the Si lifetime of the TiN-Al-10 barrier is compared to the ideal TiON-10 barrier, on a common color scale. Considering that the original unprocessed lifetime of the wafers used here was slightly above 1 ms, this direct comparison clearly reveals that the TiN-Al-10 still shows some degradation, unlike the TiON-10 barrier. The second relevant observation is that this TiN-Al-10 layer has a clearly superior barrier quality compared to the TiN-5 and TiN-10 samples shown in **Figure 1** and **Figure S1**. This is consistent with the EQE results of **Figure 4 (b)**, **Figure 6 (b)** and **Figure S2 (b)**, where the Si bottom cell is contributing with a higher current in the TiN-Al-10. In this case, the main limitation in tandem cell performance is a poor optical transmittance of this barrier layer, i.e., around 50% in the wavelength range handled by the bottom Si cell. The combination of this degradation and low transmittance helps justify the relatively low $V_{oc}$ reported for this TiN-Al-10 barrier. Therefore, in this case the TiN-Al-10 barrier achieves a similar performance by sacrificing transmittance in favor of barrier quality.

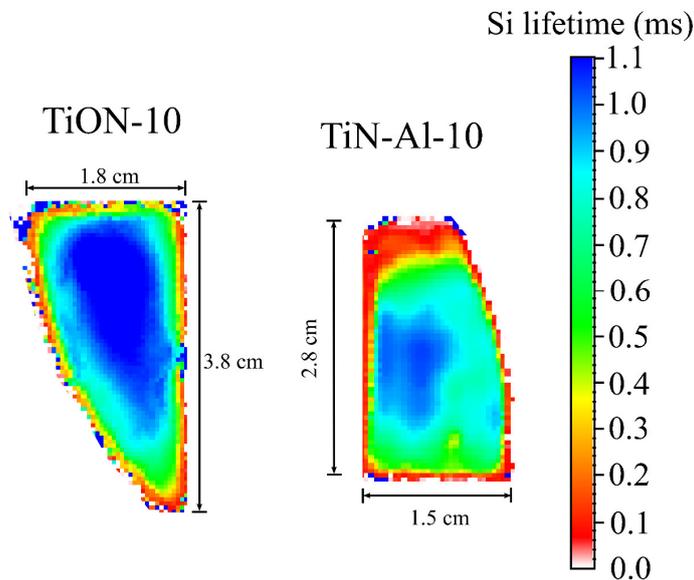

**Figure 7** – Direct comparison between the TiON-10 and TiN-Al-10 barrier layers in terms of the Si minority carrier lifetime. The lifetime scale is the same for both pieces. The annealing temperature was 575 °C.

For comparison, the transmittance spectra of all the barrier layers used in this study are shown in **Figure 8**. An additional TiN thickness value of 2.5 nm, not applied to tandem integration in this work, is added for further comparison. The region of most interest here is for wavelengths above 721 nm (below 1.72 eV), corresponding to the light the bottom Si cell would receive in an ideal bandgap-matched tandem cell, and this region is highlighted in grey, along with the CZTS bandgap energy measured in this work. In this ideal region below 1.72 eV, a reduction in TiN thickness is very effective in increasing the transmittance, and for a TiN thickness of 2.5 nm a near-ideal transmittance is achieved, similar to that of TiON-10. Notably, the transmittance is significantly superior to that of interfacial layers based on TCOs, as tested previously for CZTS/Si tandem cells[4] and bifacial kesterite cells.[18]

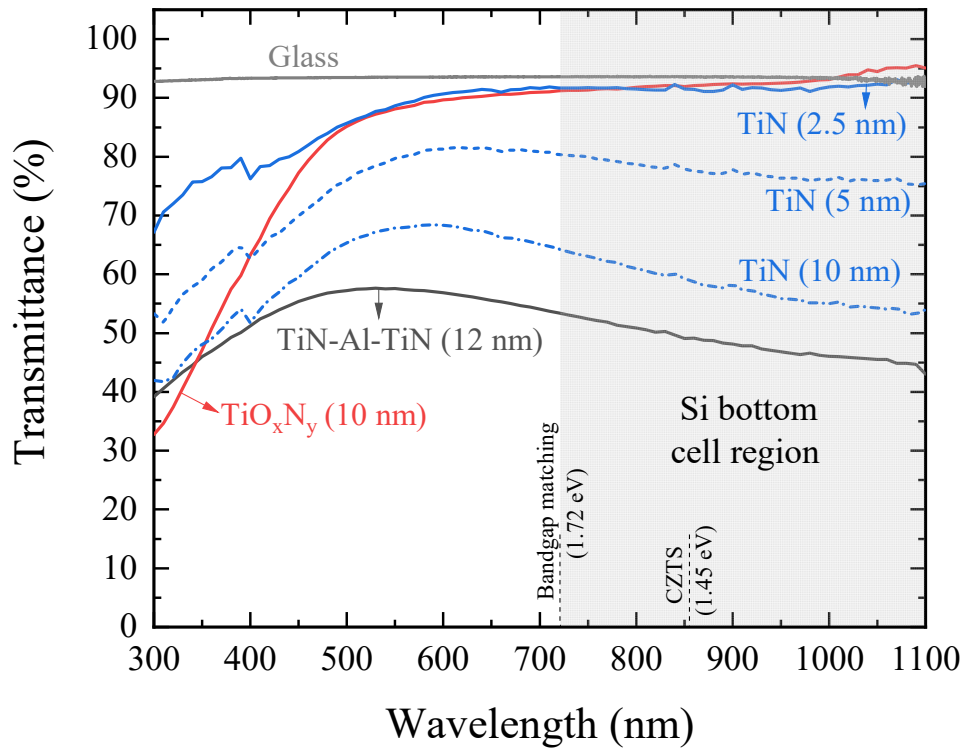

**Figure 8** – Optical transmittance of the interfacial barrier layers used in this study, as deposited on a fused silica glass substrate. For reference, the ideal bandgap matching point and the CZTS bandgap measured in this work are marked.

The fundamental trade-off, then, is that while the TiON-10 appears to be a successful barrier layer in protecting the bottom Si, it is insulating and shows a current-blocking behavior. The TiN and TiN-Al barriers, on the other hand, despite enabling a good interfacial connection, have a lower threshold for barrier failure, which does not match the necessary temperatures to produce efficient CZTS top cells. The lower the CZTS annealing temperature, the closer it is to the barrier failure threshold, as is illustrated in **Figure 9** for the TiN-5 case, for non-patterned samples. Given that the initial lifetime of the corresponding wafers was around 1 ms, the results show that this barrier failure threshold should still be much below 545 °C. On the other hand, the single-junction CZTS efficiency progressively decreases with annealing temperature (from close to 6% at 575 °C to under 4% at 545 °C, as shown in the supplementary **Figure S8**).

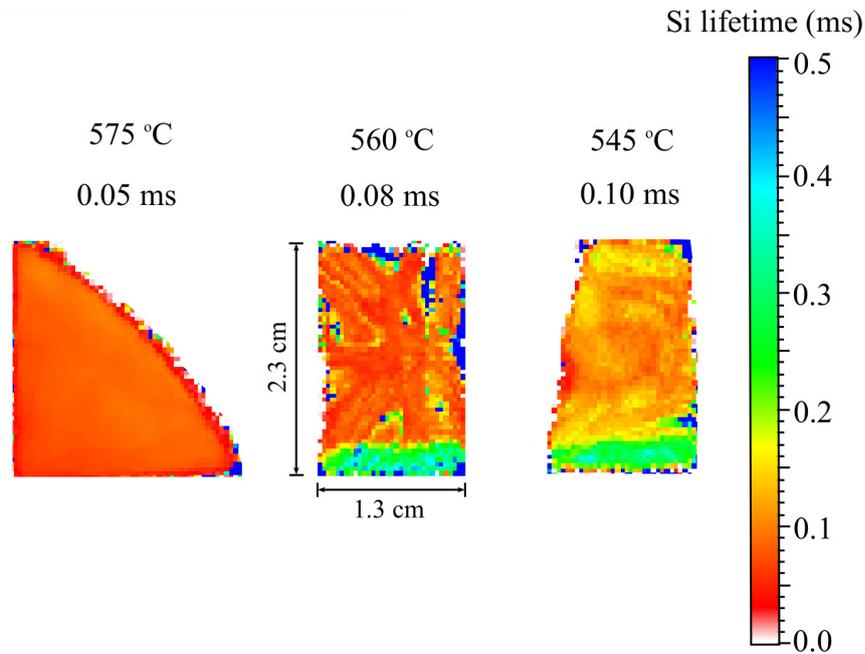

**Figure 9** – Si bottom cell lifetime comparison for CZTS annealed at 575, 560 and 545 °C, for a 5 nm TiN (TiN-5) barrier layer, without patterning. The higher lifetime region on the upper part of the wafers at 560 °C and 545 °C corresponds to an area where CZTS was not deposited due to unintentional shadowing.

In this regard, further work needs to be carried out to resolve this trade-off and find the ideal barrier layer that maximizes optical transparency, Si bottom cell protection and results in a lossless electrical interconnection. A summarized comparison between the interfacial layers used to produce monolithic CZTS/Si tandem cells is shown in **Table 2**.

**Table 2** – Comparison of the barrier layers used in monolithic CZTS/Si tandem cells. The results of this work correspond to a CZTS sulfurization temperature of 575 °C. (notes: n.a. – not available; PCE – power conversion efficiency; BSF – back surface field; The TiN transmittance values are corrected to account for the glass substrate transmittance; The Si $V_{oc}$ values were measured in small-area cleaved device wafers, and are therefore slightly underestimated)

| Interface layer | Thickness (nm) | Si Protection | Si $V_{oc}$ (mV) | Resistivity (Ωcm) | Transmittance Si region (%) | Comments | Reference |
|---|---|---|---|---|---|---|---|
| **TiON-10** | 10 | ideal | 661 | 40 | ~100 | Current blocking | This work, [3] |
| **TiN-10** | 10 | poor | 443-500 | 5 x 10$^{-3}$ | 57-69 | PCE 1.8-2.9% | This work (SI) |
| **TiN-5** | 5 | poor | 510-543 | 5 x 10$^{-3}$ | 81-86 | PCE 3-3.9% | This work |
| **TiN-2.5** | 2.5 | n.a. | n.a. | 5 x 10$^{-3}$ | >97 | Future work suggestion | This work |
| **TiN-Al-10** | 10-12 | medium | 409-450 | 4.8 x 10$^{-3}$ | 46-57 | PCE 2-2.8% | This work |
| **ZnO/MoS$_2$** | n.a. | failed | n.a | n.a. | n.a | No working device (barrier failure) | [4] |
| **ITO/MoS$_2$** | n.a. | n.a. | n.a | n.a. | n.a | No working device (adhesion issues) | [4] |
| **ZnO/FTO/MoS$_2$** | 50/650/60 | good | 545 | n.a. | 45-55 | PCE 3.5%; BSF degradation | [4] |

The results of these studies indicate that the degradation of the Si bottom cell is one of the main obstacles in achieving monolithic integration of CZTS on Si. This work shows that engineering the interfacial layer has a crucial impact on the feasibility of this monolithic integration, as it can make a difference between a full Si protection and varying degrees of degradation. Future research in alternative diffusion barrier materials could solve the trade-offs mentioned above. In this sense, for future work we suggest the possibility of using other nitride-based compounds such as tantalum nitrides, which are especially effective against Cu diffusion.[19]

Given the priority of testing the monolithic integration feasibility for the different barrier layers, in this work we kept the CZTS top cell constant as per our internal baseline. Unfortunately, our single-junction CZTS baseline is based on a disordered, and thus low bandgap kesterite, which leads to an admittedly poor bandgap matching with Si. Moreover, we note that no alkali doping (for instance through NaF) was introduced in the tandem cell at this stage, but from the work of Valentini *et al* this seems to be an improvement.[4] By solving the tandem feasibility problem, such strategies could then be employed in the future to optimize the top cell. In that case, the success of a CZTS/Si tandem cell will only be dependent on future improvements in the efficiency of CZTS-based solar cells.

Much like CZTS (and all its cationic-substituted alloys), there are several other polycrystalline materials that could be combined with Si in a tandem for solar energy conversion devices. Such materials can also involve complex multi-step synthesis approaches and often require high-temperature steps. Some examples are the high-bandgap $CuGaSe_2$ (CGS) in the photovoltaic field, and numerous high-bandgap metal-oxide photoanode materials in the photoelectrochemistry field such as $BiVO_4$[20], $WO_3$[21] and $\alpha\text{-}Fe_2O_3$[22], all of which may require, in many cases, processing temperatures above 500 °C. The feasibility of their monolithic integration with Si will then be equally dependent on the preservation of the bottom Si part of the tandem. Naturally, the results of this work are not immediately transferable, and each of these cases should be assessed individually. Nevertheless, considering the heavy contamination and degradation impact imposed by CZTS on Si (namely due to the presence of Cu and S in a high-temperature sulfurization, as we have discussed before[3]), we suggest that the monolithic integration of materials such as those mentioned above could also be feasible and promising, despite very little research in that regard. In fact, a very promising report of a monolithic CGS/Si tandem solar cell already exists, where almost no Si degradation occurred, in spite of the high-temperature step involved in the CGS fabrication.[23] A tandem architecture based on an interfacial diffusion barrier, such as the one proposed in this work, could then be a potentially interesting configuration for future research.

Lastly, it is also worth mentioning that the resilience of the Si bottom cell is also dependent on the architecture choice of the Si cell itself. In this work, we used a thermally resilient TOPCon Si configuration, and the properties of the different Si layers were kept fixed, changing only the TiN-based layers. However, it is likely that further improvements could be made in the bottom Si cell structures in order to increase its resistance against CZTS processing. One example would be the n+ and p+PolySi selective contacts, where a change in thickness or doping density could have a meaningful effect in protecting the n-Si bulk. Our current research is focusing on these possibilities.

**Experimental Methods**

The Si bottom cells used in this work consist of a double-sided tunnel oxide passivated contact (TOPCon) structure, where $SiO_2$ tunnel oxide (TO) and heavily doped polysilicon layers (p+PolySi and n+PolySi) were grown on each side of a monocrystalline n-Si substrate. In order to improve the surface passivation quality of the top and bottom interfaces, a hydrogenation step was included by depositing a 75 nm sacrificial SiN:H layer on both sides of the wafer and performing a hydrogen drive-in annealing at 400 °C. Before the growth of the TiN and CZTS layers, the SiN:H layer on the n+PolySi side was removed using a buffered HF solution. The SiN:H layer on the p+PolySi side was only removed after the complete processing of the CZTS top cell, to enhance the protection of the backside of the wafer during the different processing steps (more details are given below). The bottom Si cell configuration used for the subsequent CZTS processing, including the TiN barrier layers, was SiN:H/p+PolySi/TO/n-Si/TO/n+PolySi/TiN (sequence starting from the backside). Details on the fabrication of the bottom cell structures were mentioned in previous work.[3] The TiN and TiON barrier layers were deposited in a Picosun Plasma-Enhanced ALD (PEALD) system using TiCl$_4$ and NH$_3$ precursors at 400 °C. The oxygen incorporation in TiON was achieved by running the TiN ALD recipe without prior chamber passivation (i.e. a sequence of dummy TiN depositions), thereby allowing a higher oxygen background level. The TiN-Al-TiN barrier layer was deposited by sequential sputtering of a 5 nm TiN layer, a 2 nm Al layer and a second 5 nm TiN layer, without breaking vacuum. The resulting structure was then annealed in air at 350 °C for 30 minutes.

For the experiments with surface patterning, the front SiN layer (used for the hydrogenation process) was patterned by means of photolithography and wet etching in buffered HF solution for 5 min. The surface was then cleaned thoroughly in RCA1 and RCA2 solutions prior to the barrier layer deposition.

The CZTS absorber layers for the top cell of the tandem were fabricated by a two-step process. First, Cu, ZnS and SnS precursors were co-sputtered directly onto the TiN/Si bottom cell stack mentioned above, and subsequently annealed in a graphite box with a sulfur and tin sulfide-containing atmosphere, using 50 mg of S pellets and 5 mg of Sn powder and N$_2$ gas. The main CZTS annealing temperature used in this work was 575 °C, except in one set of experiments, where temperatures of 545 and 560 °C were used to study the TiN barrier failure threshold by measuring the resulting Si lifetime after annealing. The heating rate was 20 °C/min and the dwelling time was 45 min. According to cross-section secondary electron images (not shown here), the final thickness of the CZTS absorbers varied between 275-350 nm. Following the CZTS annealing, a 50 nm CdS layer was deposited by chemical bath deposition to form the top cell p-n heterojunction. A 50 nm intrinsic ZnO (i-ZnO) layer was deposited by reactive sputtering, followed by a 350 nm Al-doped ZnO (AZO) layer deposited by Ar sputtering to form the top electrode of the tandem cell. Finally, the SiN:H sacrificial layer on the back side was removed in a buffered HF solution, and 500 nm Ag was evaporated to form the bottom electrode of the tandem cell. In order to protect the top layers during the buffered HF etch, a AZ5214E photoresist was spin coated on the AZO top electrode, and subsequently removed using acetone. A post-fabrication thermal treatment was applied to the full tandem solar cells, by annealing at 275 °C for 8 min in a $N_2$ atmosphere, in order to improve the properties of the CZTS/CdS heterojunction interface.[24]

The effective minority carrier lifetime maps of the Si bottom cells were measured by the microwave detected photoconductance decay method (μ-PCD) in a steady-state configuration at 1-sun illumination using an MDP lifetime scanner from Freiberg Instruments. The Si lifetime was measured immediately before and after the CZTS processing steps, to evaluate the protecting effect of the barrier layers. In general, the lifetime was measured from the back side (SiN:H/p+PolySi side). However, two extra verification procedures were done to insure the reliability of the lifetime measurements. In one, the lifetime values were measured before the Ag deposition (i.e. full top cell processing included). In the other, the SiN, TiN and CZTS layers were chemically removed after the CZTS annealing step, using a mixture of $H_2O_2:4H_2SO_4$ (piranha) and RCA1 solutions, followed by a dilute HF dip. Afterwards, the lifetime was measured on the remaining TOPCon Si stack (PolySi/TO/n-Si/TO/PolySi). The three different measurement configurations yielded similar results.

For the X-ray Photoelectron Spectroscopy measurements, the surface of the samples was slightly sputtered-off (*in situ*) by means of $Ar^+$ ions with 500 eV and 60 s for each cycle, to remove the existing Ti native oxide and possible carbon contamination before acquiring the measurements. An Al $K_α$ source was used, and the spot size was 400 μm.

To further compare the TiN and TiON barrier layer quality, a separate Cu diffusion experiment was conducted. For this purpose, 100 nm of metallic Cu layers were sputtered on TOPCon samples with TiN and TiON barrier layers, and annealed at 550 °C in vacuum ($1 \times 10^{-6}$ mbar) for a series of different annealing times (15, 30, 45 and 60 minutes). After annealing, the Cu and TiN layers were chemically removed using a mixture of $H_2O_2$:$4H_2SO_4$ (piranha) and RCA1 solutions, followed by a dilute HF dip. Then, the quantitative Cu depth profiles were measured by secondary ion mass spectrometry (SIMS). The SIMS depth profiles were obtained from a Cameca IMS-7f microprobe. A 10 keV $O^{2+}$ primary beam was mainly utilized, and rastered over $150 \times 150$ μm$^2$, and the positive ions were collected from a circular area with a diameter of 33 μm. The Cu concentration was obtained by measuring an implanted reference sample. The crater depths were measured by a Dektak 8 stylus profilometer, and a constant sputter erosion rate was assumed for the depth calculation.

The J-V characteristic curves of the solar cells were measured at near Standard Test Conditions (STC: 1000 W/m$^2$, AM 1.5 and 25 °C). A Newport class ABA steady state solar simulator was used. The irradiance was measured with a $2 \times 2$ cm$^2$ Mono-Si reference cell from ReRa certified at STC by the Nijmegen PV measurement facility. The temperature was kept at $25 \pm 3$ °C as measured by a temperature probe on the contact plate. The acquisition was done with 2 ms between points, using a 4 wire measurement probe, from reverse to forward voltage. The $V_{oc}$ produced by the bottom Si cell of the tandem was measured by mechanically scratching a small area of the surface of the tandem, exposing the TiN surface (which, due to its hardness, does not scratch as easily as the top cell layers above it). Then, an additional J-V curve was measured by contacting the exposed TiN area with the negative probe, and the $V_{oc}$ of the result curve could be directly identified as the bottom cell $V_{oc}$ for the tandem.

The external quantum efficiency (EQE) of the tandem cell was measured using a QEXL setup (PV Measurements) equipped with a grating monochromator, adjustable bias voltage, and a bias spectrum. The CZTS top cell was measured with light-biasing of the bottom Si cell using a high-pass filter at 900 nm, and the Si bottom cell was measured with light-biasing of the top CZTS cell using a band-pass filter from 400-500 nm.

ASSOCIATED CONTENT

**Supporting Information**

The supporting information includes further details on the lifetime degradation effects of the Si bottom cell for different barrier layer configurations, as well as additional characterization on the tandem cells produced in this work.

AUTHOR INFORMATION

**ORCID**

Filipe Martinho: 0000-0002-0616-3260

Alireza Hajijafarassar: 0000-0003-2716-7639

Mungunshagai Gansukh: 0000-0002-2181-9862

Stela Canulescu: 0000-0003-3786-2598

Andrea Crovetto: 0000-0003-1499-8740

Jørgen Schou: 0000-0002-8647-2679

Ole Hansen: 0000-0002-6090-8323

**Notes**

The authors declare no competing financial interest.

ACKNOWLEDGMENT

This work was supported by a grant from the Innovation Fund Denmark (grant number 6154-00008A). F.M. would like to thank A.A. Santamaria Lancia for the support on the J-V measurements.

Photoelectron Spectroscopy Study. *J. Appl. Phys.* **1992**, *72* (7), 3072–3079. https://doi.org/10.1063/1.351465.

(13) Donolato, C. A Reciprocity Theorem for Charge Collection. *Appl. Phys. Lett.* **1985**, *46* (3), 270–272. https://doi.org/10.1063/1.95654.

(14) Donolato, C. Reciprocity Theorem for Charge Collection by a Surface with Finite Collection Velocity: Application to Grain Boundaries. *J. Appl. Phys.* **1994**, *76* (2), 959–966. https://doi.org/10.1063/1.357774.

(15) Brody, J. Review and Comparison of Equations Relating Bulk Lifetime and Surface Recombination Velocity to Effective Lifetime Measured under Flash Lamp Illumination. *Sol. Energy Mater. Sol. Cells* **2003**, *77* (3), 293–301. https://doi.org/10.1016/S0927-0248(02)00350-1.

(16) Luke, K. L.; Cheng, L. Analysis of the Interaction of a Laser Pulse with a Silicon Wafer: Determination of Bulk Lifetime and Surface Recombination Velocity. *J. Appl. Phys.* **1987**, *61* (6), 2282–2293. https://doi.org/10.1063/1.337938.

(17) Green, M. A.; Keevers, M. J. Optical Properties of Intrinsic Silicon at 300 K. *Prog. Photovoltaics Res. Appl.* **1995**, *3* (3), 189–192. https://doi.org/10.1002/pip.4670030303.

(18) Espindola-Rodriguez, M.; Sylla, D.; Sánchez, Y.; Oliva, F.; Grini, S.; Neuschitzer, M.; Vines, L.; Izquierdo-Roca, V.; Saucedo, E.; Placidi, M. Bifacial Kesterite Solar Cells on FTO Substrates. *ACS Sustain. Chem. Eng.* **2017**, *5* (12), 11516–11524. https://doi.org/10.1021/acssuschemeng.7b02797.

(19) Min, K.-H. Comparative Study of Tantalum and Tantalum Nitrides ($Ta_2N$ and $TaN$) as a Diffusion Barrier for Cu Metallization. *J. Vac. Sci. Technol. B Microelectron. Nanom. Struct.* **1996**, *14* (5), 3263. https://doi.org/10.1116/1.588818.

**TOC GRAPHICS**

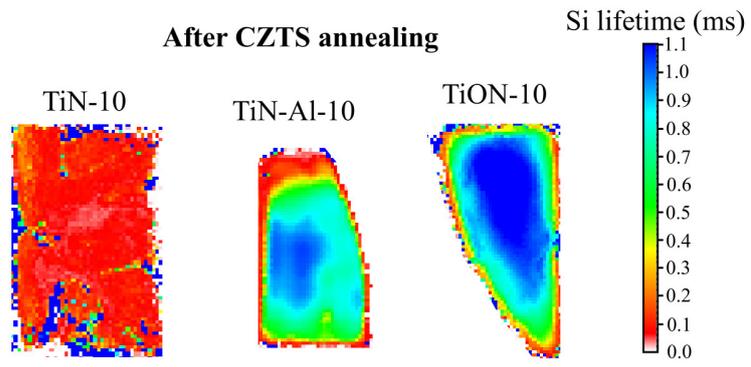